\documentclass[submission, Phys]{SciPost}

\hypersetup{
    colorlinks,
    linkcolor={red!50!black},
    citecolor={blue!50!black},
    urlcolor={blue!80!black}
}

\usepackage[bitstream-charter]{mathdesign}
\DeclareSymbolFont{usualmathcal}{OMS}{cmsy}{m}{n}
\DeclareSymbolFontAlphabet{\mathcal}{usualmathcal}

\begin{document}

\begin{center}{\Large \textbf{ Tau Data-Based Evaluations of the hadronic vacuum polarization contribution to the muon g-2
\\
}}\end{center}

\begin{center}
Pere Masjuan\textsuperscript{1,2},
Alejandro Miranda\textsuperscript{2} and
Pablo Roig\textsuperscript{3$\star$}
\end{center}

\begin{center}
{\bf 1} Grup de F\'isica Te\`orica, Departament de F\'isica, Universitat Aut\`onoma de Barcelona, 08193
Bellaterra (Barcelona), Spain.
\\
{\bf 2} Institut de F\'isica d’Altes Energies (IFAE) and The Barcelona Institute of Science and
Technology (BIST), Campus UAB, 08193 Bellaterra (Barcelona), Spain.
\\
{\bf 3} Departamento de F\'isica, Centro de Investigaci\'on y de Estudios Avanzados del Instituto
Polit\'ecnico Nacional, Apdo. Postal 14-740, 07000 Ciudad de M\'exico, M\'exico.
\\[\baselineskip]
$\star$ \href{mailto:pablo.roig@cinvestav.mx}{\small pablo.roig@cinvestav.mx}
\end{center}

\begin{center}
\date{}
\end{center}


\definecolor{palegray}{gray}{0.95}
\begin{center}
\colorbox{palegray}{
  \begin{tabular}{rr}
  \begin{minipage}{0.1\textwidth}
    \includegraphics[width=30mm]{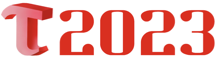}
  \end{minipage}
  &
  \begin{minipage}{0.81\textwidth}
    \begin{center}
    {\it The 17th International Workshop on Tau Lepton Physics}\\
    {\it Louisville, USA, 4-8 December 2023} \\
    \doi{10.21468/SciPostPhysProc.?}\\
    \end{center}
  \end{minipage}
\end{tabular}
}
\end{center}

\section*{Abstract}
{\boldmath\textbf{
We review the $\tau$ data-driven computation of Euclidean windows for the hadronic vacuum polarization contribution to the muon anomalous magnetic moment ($a_\mu$), which agree with the lattice results, making the difference of the $e^+e^-$ data-driven methods with them more intriguing. This conundrum needs to be solved by next year, when the final FNAL $a_\mu$ measurement will be published, in order to extract firm conclusions on possible new physics effects.}}


\section{Introduction}
\label{sec:intro}
The first two publications from the FNAL Muon g-2 Collaboration \cite{Muong-2:2021ojo, Muong-2:2023cdq}, together with the BNL final result \cite{Muong-2:2006rrc} yield the current experimental world average
\begin{equation}\label{eq.g-2Exp}
a_\mu^{\mathrm{exp}}=116 592 059(22)\times10^{-11}\,,
\end{equation}
with an impressive accuracy of $0.19$ppm.\\

Although there is intense work, within the Muon g-2 Theory Initiative, to update the corresponding SM prediction, its recommended value is still the one in the White Paper \cite{Aoyama:2020ynm}
\begin{equation}\label{eq.g-2SM}
a_\mu^{\mathrm{SM}}=116 591 810(43)\times10^{-11}\,,
\end{equation}
which is based on Refs.~\cite{Davier:2017zfy,Keshavarzi:2018mgv,Colangelo:2018mtw,Hoferichter:2019mqg,Davier:2019can,Keshavarzi:2019abf,Kurz:2014wya,FermilabLattice:2017wgj,Budapest-Marseille-Wuppertal:2017okr,RBC:2018dos,Giusti:2019xct,Shintani:2019wai,FermilabLattice:2019ugu,Gerardin:2019rua,Aubin:2019usy,Giusti:2019hkz,Melnikov:2003xd,Masjuan:2017tvw,Colangelo:2017fiz,Hoferichter:2018kwz,Gerardin:2019vio,Bijnens:2019ghy,Colangelo:2019uex,Pauk:2014rta,Danilkin:2016hnh,Jegerlehner:2017gek,Knecht:2018sci,Eichmann:2019bqf,Roig:2019reh,Colangelo:2014qya,Blum:2019ugy,Aoyama:2012wk,Aoyama:2019ryr,Czarnecki:2002nt,Gnendiger:2013pva} (see also, e.g. \cite{Colangelo:2022jxc,Miranda:2020wdg,Masjuan:2020jsf,Ludtke:2020moa,Bijnens:2021jqo,Cappiello:2021vzi,Chao:2021tvp,Colangelo:2021moe,Colangelo:2021nkr,Danilkin:2021icn,Hoferichter:2021wyj,Leutgeb:2021mpu,Miramontes:2021exi,Giusti:2021dvd,Biloshytskyi:2022ets,Colangelo:2022prz,Bijnens:2022itw,Boito:2022rkw,Boito:2022dry,Stamen:2022uqh,Masjuan:2023qsp,RBC:2023pvn,Ludtke:2023hvz,Davier:2023hhn,Wang:2023njt,Ce:2022kxy,ExtendedTwistedMass:2022jpw,Colangelo:2022vok,Benton:2023dci,Lehner:2020crt,Wang:2022lkq,Aubin:2022hgm,FermilabLatticeHPQCD:2023jof,Davier:2023cyp,Davier:2023fpl,Hoferichter:2023sli,Bijnens:2020xnl,Chao:2020kwq,Colangelo:2020lcg,Hoid:2020xjs,Miranda:2021lhb,Escribano:2023seb,Roig:2014uja,Guevara:2018rhj,Raya:2019dnh}). The $5.1\sigma$ deviation between Eqs.~(\ref{eq.g-2Exp}) and (\ref{eq.g-2SM}) needs to be taken cautiously, as sketched below.\\

The BMW collaboration \cite{Borsanyi:2020mff} achieved a very precise lattice QCD evaluation of the hadronic vacuum polarization contribution (HVP) which dominates the error of the SM prediction, and challenged the WP number four years ago. According to this result, the discrepancy with respect to Eq.~(\ref{eq.g-2Exp}) would be reduced to $1.5\sigma$. There is not yet another lattice QCD computation of similar accuracy which can be compared with the BMW number. Since Ref.~\cite{Colangelo:2022vok}, it has become standard to compare data-driven and lattice QCD evaluations of $a_\mu^\text{HVP}$ in three different windows in Euclidean time, which we pioneered using $\tau$ decay input in Ref.~\cite{Masjuan:2023qsp} (see also \cite{Davier:2023fpl}), on which this contribution is fundamentally based. Remarkably, in the so-called intermediate window (to be introduced and discussed below), $\tau$-based evaluations \cite{Masjuan:2023qsp, Davier:2023fpl} and the Mainz CLS \cite{Ce:2022kxy}, ETMC  \cite{ExtendedTwistedMass:2022jpw} and RBC/UKQCD \cite{RBC:2023pvn} collaborations agree closely with BMW \cite{Borsanyi:2020mff}, with competing precision.\\

At the beginning of 2023, the CMD-3 Collaboration \cite{CMD-3:2023alj,CMD-3:2023rfe} released their measurement of $\sigma(e^+e^-\to\pi^+\pi^-)$, which disagreed substantially with previous data \cite{CMD-2:2003gqi,CMD-2:2005mvb,CLEO:2005tiu,Aulchenko:2006dxz,CMD-2:2006gxt,Achasov:2006vp,KLOE:2010qei,KLOE:2012anl,KLOE-2:2017fda,BaBar:2012bdw,BESIII:2015equ,SND:2020nwa} though was in line with the BMW result. $a_\mu^\text{HVP}$ based on CMD-3 alone would yield agreement between the SM prediction and Eq.~(\ref{eq.g-2Exp}) within one sigma.\\

In this puzzling situation we insist \cite{Miranda:2020wdg, Masjuan:2023qsp} on the value of data-driven determinations of $a_\mu^\text{HVP}$ using $\tau$ input for the dominant di-pion channel, giving $\sim73(85)\%$ of the whole value(uncertainty). An additional independent determination of $a_\mu^\text{HVP}$ will come soon from the space-like measurement in MUonE \cite{Banerjee:2020tdt, Abbiendi:2022oks} and will help to settle this question. After the end of the FNAL muon g-2 experiment, the J-PARC one will start operating, with the added value of being affected by completely different systematics \cite{Iinuma:2011zz}. The community hopes that this effort will further strength $a_\mu^{\mathrm{exp}}$ and increase its precision, motivating to continue the improvement of $a_\mu^{\mathrm{SM}}$ beyond 2025.

\section{\texorpdfstring{$\boldsymbol{a_\mu^\text{HVP}}$}{Lg} in the SM}
The $e^+e^-$ data-driven evaluation of the leading order HVP contribution to the muon g-2 is computed from
\begin{equation}
a_\mu^\text{HVP, LO}=\frac{1}{4\pi^3}\int_{s_\text{thr}}^{\infty}\mathrm{d}s\, K(s)\, \sigma^0_ {e^+e^-\to\mathrm{hadrons}(\gamma)}(s)\,,
\end{equation}
which uses the measurement of the bare hadronic cross-section (specified by the upper-index $^0$). We note that both the $1/s$ behaviour of the cross-section (away from resonance peaks) and the $K(s)$ kernel enhance the low-energy contributions, making the precise evaluation of the $\pi^+\pi^-$ cut critical. Currently, the discrepancy between the BaBar \cite{BaBar:2012bdw} and KLOE \cite{KLOE-2:2017fda} results has been very much aggravated by the CMD-3 data \cite{CMD-3:2023alj}, jeopardizing this data-driven evaluation at the level of precision currently needed.\\

In this context, it is worth to resurrect the corresponding evaluation using $\tau$ data, that was once very useful in combination with the $e^+e^-$ measurements to improve the accuracy of the SM prediction \cite{Alemany:1997tn}, and has been used fruitfully since then \cite{Narison:2001jt,Cirigliano:2001er,Cirigliano:2002pv,Davier:2002dy,Davier:2003pw,Maltman:2005yk,Maltman:2005qq,Davier:2010fmf,Davier:2010nc,Benayoun:2012etq,Davier:2013sfa,Bruno:2018ono,Narison:2023srj,Esparza-Arellano:2023dps}.\\

This requires isospin-breaking (IB) corrections to relate the measured di-pion $\tau$ decay spectrum to the needed bare cross-section

\begin{equation}
\sigma^0_{\pi\pi(\gamma)}=\left[\frac{K_\sigma(s)}{K_\Gamma(s)}\frac{\mathrm{d}\Gamma_{\pi\pi(\gamma)}}{\mathrm{d}s}\right]\frac{R_\text{IB}(s)}{S_\text{EW}}\,,
\end{equation}
where the first factor includes kinematic dependencies and global constants and $S_\text{EW}$ corresponds to the universal short-distance radiative correction (see its updated value in \cite{Masjuan:2023qsp}). The most difficult corrections to evaluate are inside $R_\text{IB}(s)$ \cite{Cirigliano:2001er,Davier:2010fmf}
\begin{equation}
R_\text{IB}(s)=\frac{\mathrm{FSR}(s)}{G_\text{EM}(s)}\frac{\beta_{\pi^+\pi^-}^3(s)}{\beta_{\pi^-\pi^0}^3(s)}\Bigg| \frac{F_V(s)}{f_+(s)}\Bigg|^2\,,
\end{equation}
where the easy part (for the needed precision) corresponds to the phase-space correction coming from the ratio of $\beta$ functions and the final state radiative correction, FSR. It is much more difficult to quantify the long-distance process-dependent electromagnetic corrections entering the $G_\text{EM}(s)$ \cite{Cirigliano:2001er} and the difference between the neutral ($F_V$) and charged ($f_+$) di-pion form factors.\\

The $S_\text{EW}$ contribution gives $\Delta a_\mu=-119.6\times10^{-11}$, consistent with earlier determinations and with a negligible error. The PS correction induces $\Delta a_\mu=-74.5\times10^{-11}$, as in previous works. The FSR correction yields $\Delta a_\mu=+45.5(4.6)\times10^{-11}$, in accord with \cite{Davier:2010fmf}.\\

We evaluate \cite{Miranda:2020wdg} the structure-dependent IB corrections using Chiral Perturbation Theory \cite{Weinberg:1978kz,Gasser:1983yg,Gasser:1984gg} extended with Resonance fields and accounting for short-distance QCD constraints \cite{Ecker:1988te,Ecker:1989yg,Cirigliano:2006hb,Kampf:2011ty,Roig:2013baa}, using a large-$N_C$ expansion \cite{tHooft:1973alw}.\\

We find that IB in the form factors induces the shift $+40.9(48.9)\times10^{-11}$ or $+77.6(24.0)\times10^{-11}$, depending on the input that we use. These results agree with earlier evaluations \cite{Cirigliano:2001er,Davier:2010fmf}, all with sizable uncertainties. For the $G_\text{EM}$ effect, we obtain $(-15.9^{+5.7}_{-16.0})\times10^{-11}$, which agrees with Refs. \cite{Cirigliano:2001er,Davier:2010fmf}.\\

\begin{figure}[ht]
    \centering
    \includegraphics[width=8cm]{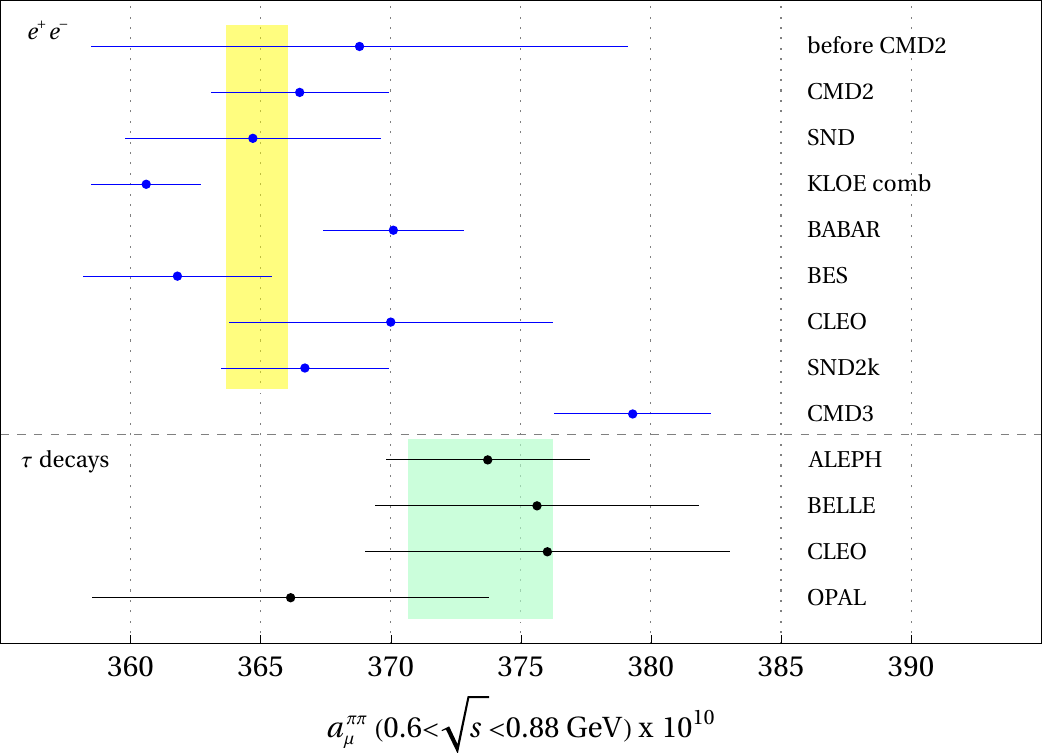}
    \caption{The $\pi\pi(\gamma)$ contribution to $a_\mu^{\text{HVP, LO}}$ around the $\rho$ peak, obtained from the $e^+e^-\to\pi^+\pi^-(\gamma)$ cross section (top) and di-pion $\tau$ decays (bottom).}
    \label{fig:amu_eetau}
\end{figure}

Fig.~\ref{fig:amu_eetau} shows the $\pi\pi$ contribution to $a_\mu^\text{HVP, LO}$ around the $\rho$ peak, found using either $\sigma(e^+e^-\to\mathrm{hadrons})$ (top, with average -without the CMD-3 point- in yellow) or the $\tau^-\to\pi^-\pi^0\nu_\tau$ spectrum (bottom, with mean in green that agrees with CMD-3). Tau data yields a $\sim10\times10^{-10}$ larger value.

\begin{figure}[ht]
    \centering
    \includegraphics[width=8cm]{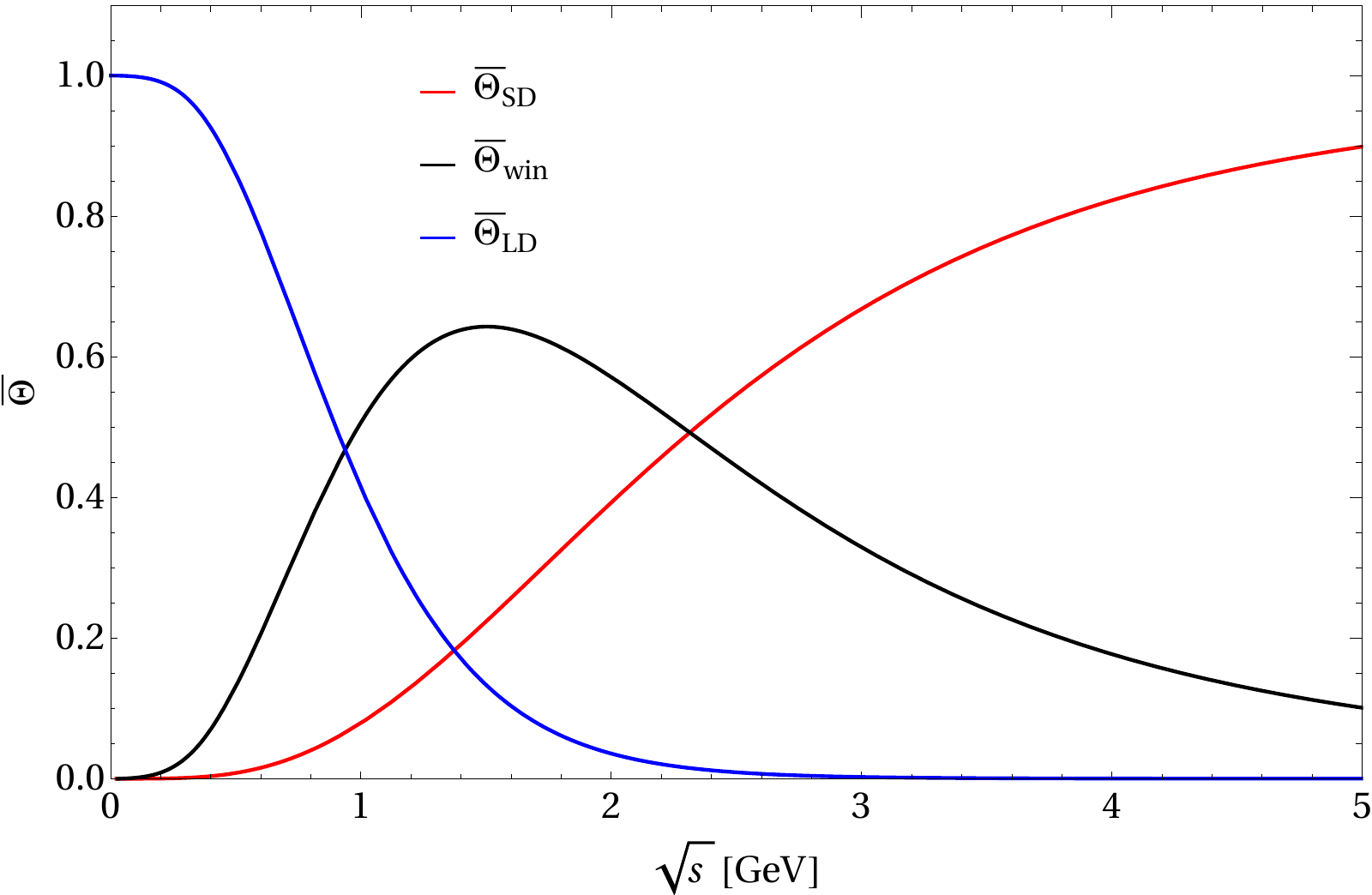}
    \caption{Normalized weights of the short-distance ($SD$), intermediate ($win$, also $int$) and long-distance ($LD$) windows in $\sqrt{s}$ \cite{Colangelo:2022vok}.}
    \label{fig:windowsE}
\end{figure}

In our recent paper \cite{Masjuan:2023qsp} we translate these results into the window quantities introduced in Ref.~\cite{RBC:2018dos}, as used in \cite{Colangelo:2022vok}. The corresponding weights in Euclidean time or energy (the latter reproduced here as Fig.~\ref{fig:windowsE}) are shown in Fig. 1 of \cite{Colangelo:2022vok}. It is important to clarify that the names can be misleading, as the intermediate window (of simply 'the window') receives significant contributions from both low and high energies. Similarly, the short-distance window is still mildly sensitive to hadronization, and the tail of the long-distance window enters the perturbative regime. This observation needs to be taken into account when interpreting results in the different windows, with respect to the expected theory uncertainty. It is also crucial that the contributions of these windows to $a_\mu^\text{HVP}$ scale as $\sim1:10:25$, respectively, so that the relative accuracy needed varies substantially between them.\\

\begin{figure}[ht!]
    \centering
    \includegraphics[width=8.2cm]{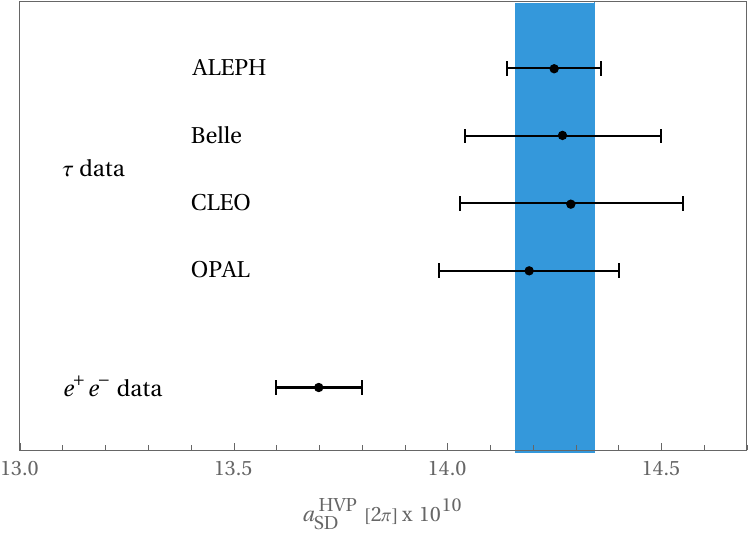}
    \includegraphics[width=8.0cm]{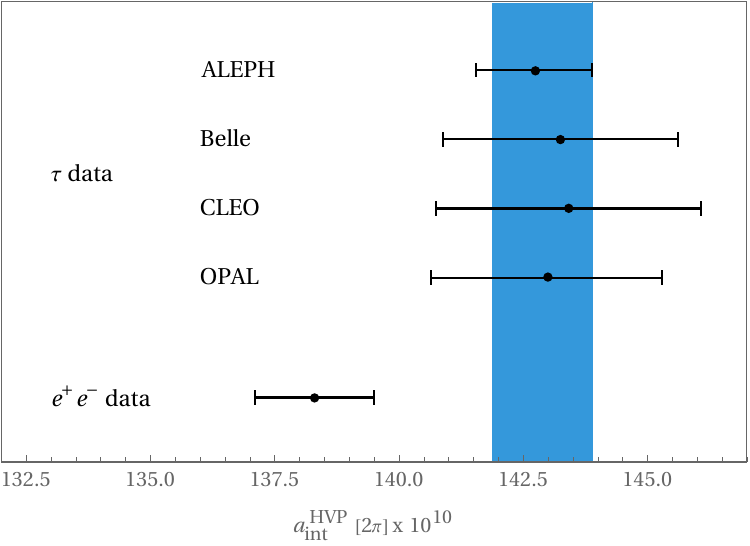}
    \includegraphics[width=8.2cm]{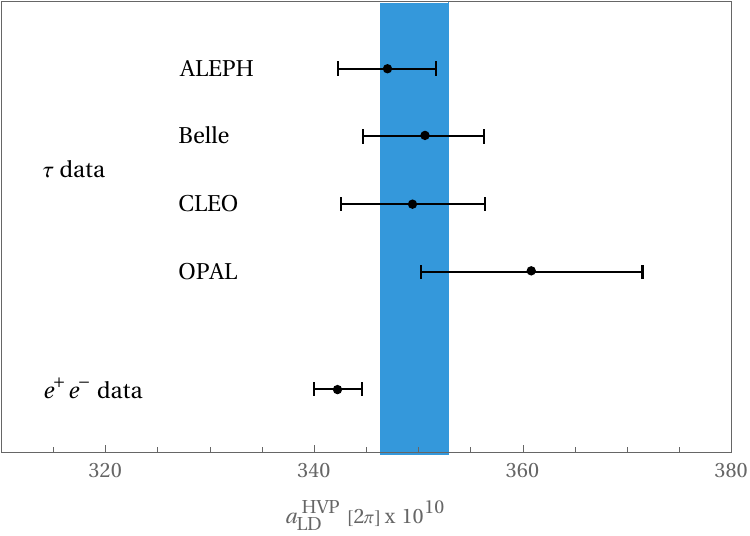}
    \caption{Window quantities ($SD$ top, $int$ medium, and $LD$ bottom) for the $2\pi$ contribution below $1.0\,\text{GeV}$ to $a_\mu^{\mathrm{HVP}}$ according to our reference results. 
 The blue band displays the mean from $\tau$ data, with the $e^+e^-$ result from \cite{Colangelo:2022vok}. }
    \label{fig:ChPTOp4}
\end{figure}

 Our main results for the three different window contributions to $a_\mu^{\mathrm{HVP}}$ are plotted in Fig.~\ref{fig:ChPTOp4}, where the consistency among the different $\tau$ measurements is evident. In the $SD$ and $int$ windows, $e^+e^-$ (from Ref.~\cite{Colangelo:2022vok}) and $\tau$ data-based values are at odds.\\

\begin{figure}[ht!]
    \centering
    \includegraphics[width=8.3cm]{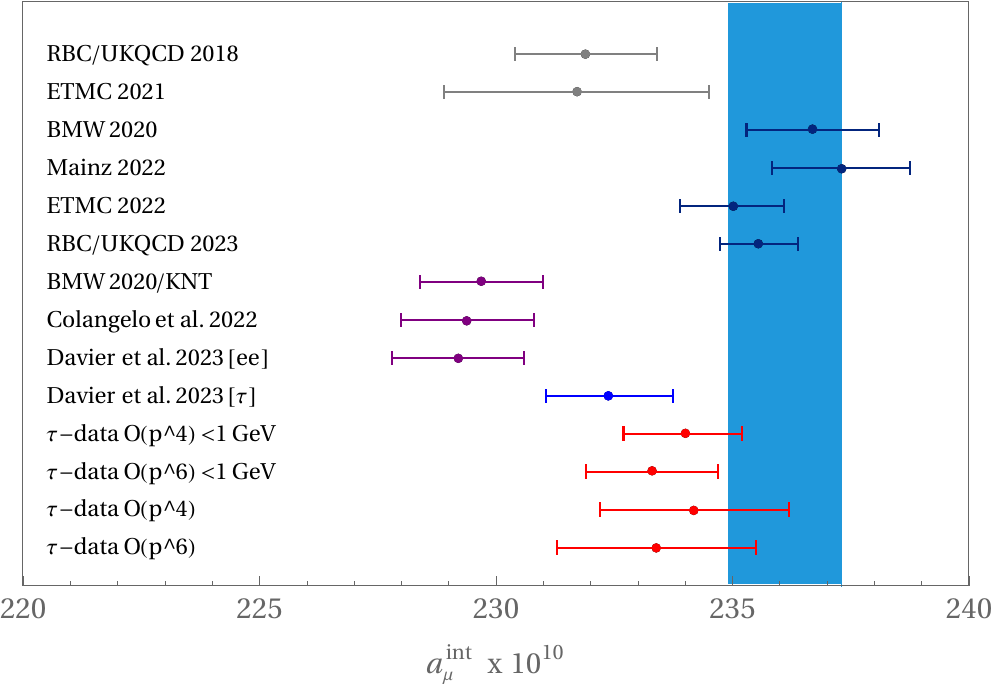}
    \caption{Comparison of the total intermediate window contribution to $a_{\mu}^{\text{HVP, LO}}$ according to lattice QCD, $e^+e^-$ and $\tau$ data-driven evaluations. The blue band corresponds to the weighted average of the lattice results excluding RBC/UKQCD 2018~\cite{RBC:2018dos} and ETMC 2021~\cite{Giusti:2021dvd} (superseded by refs.~\cite{RBC:2023pvn} and \cite{ExtendedTwistedMass:2022jpw}, respectively).}
    \label{fig:lattice_results}
\end{figure}

Our $\tau$-based $\pi\pi$ contribution to $a_\mu^\text{HVP, LO}$ is supplemented with that from the remaining modes in order to compare it directly with the full evaluations. For this we considered two approaches, as explained in detail in Ref.~\cite{Masjuan:2023qsp}, and took the difference between them as the uncertainty associated to this procedure. In this way, we found the results shown in Fig.~\ref{fig:lattice_results}. There is a clear trend of $\tau$-based evaluations agreeing with the lattice results, while the $e^+e^-$ ones differ markedly with both.

\section{Conclusion}
We have reviewed our $\tau$-based evaluation of the dominant di-pion contribution to $a_\mu^\text{HVP, LO}$ \cite{Miranda:2020wdg, Masjuan:2023qsp}. Interestingly, it agrees nicely with the lattice QCD computations in the so-called intermediate window (as well as in the other ones), with the $e^+e^-$-based results differing clearly with both. In light of these findings and the CMD-3 data (which is compatible with the lattice and $\tau$-based results), the consistency of the different $e^+e^-\to\pi^-\pi^-$ data seem puzzling, and further work appears needed to clarify this situation. This would be extremely important for interpreting the final FNAL measurement of $a_\mu$, concluding on possible new physics contributions through its comparison with the SM prediction, which is not clear enough at the moment.
\section*{Acknowledgements}
It is our pleasure to thank Swagato and his crew for this excellent workshop.
\paragraph{Funding information}
 The work of P. M. has been supported by the European
Union’s Horizon 2020 Research and Innovation Programme under grant 824093 (H2020-INFRAIA-2018-1), the Ministerio de Ciencia e Innovaci\'on under grant PID2020-112965GB-I00, and by the Secretaria d’Universitats i Recerca del Departament d’Empresa i Coneixement de la Generalitat de Catalunya under grant 2021 SGR 00649. IFAE is partially funded by the CERCA program of
the Generalitat de Catalunya. A. M. is also supported by MICINN with funding from European Union NextGenerationEU (PRTR-C17.I1) and by Generalitat de Catalunya. P. R. thanks partial funding from Conahcyt (M\'exico).






\nolinenumbers

\end{document}